\documentclass[3p,preprint]{elsarticle}
\usepackage[utf8]{inputenc}
\usepackage{graphicx}
\usepackage{dcolumn}
\usepackage{bm}
\journal{Physica E}

\begin{document}


\title{Effects of Fermi velocity engineering in magnetic graphene superlattices}

\author[UFRPE]{Ícaro S. F. Bezerra}

\author[UFRPE,INT]{Jonas R. F. Lima}%
\address[UFRPE]{Departamento de F\'{\i}sica, Universidade Federal Rural de Pernambuco, 52171-900, Recife, PE, Brazil}

\address[INT]{Institute of Nanotechnology, Karlsruhe Institute of Technology, D-76021 Karlsruhe, Germany}

\date{\today}

\begin{abstract}
In this work we investigate theoretically the influence of a Fermi velocity modulation in the electronic and transport properties of magnetic graphene superlattices. We solve the effective Dirac equation for graphene with a position dependent vector potential and Fermi velocity and use the transfer matrix method to obtain the transmission coefficient for the finite cases and the dispersion relation for a periodic superlattice. Our results reveals that the Fermi velocity modulation can control the resonance peaks of the transmittance and also works as a switch, turning on/off the transmission through the magnetic barriers. The results obtained here can be used for the fabrication of graphene-based electronic devices.

\end{abstract}

\maketitle


\section{Introduction}

Since its first experimental realization \cite{Novoselov}, graphene has been considered a promising material for the fabrication of electronic devices, due, for instance, to its extremely high carrier mobility  \cite{RevModPhys.81.109} and long-range ballistic transport at room temperature \cite{Berger1191}, which exceed those of conventional semiconductors. However, graphene has a massless Dirac spectrum at low-energies and the Klein tunneling prevents the charge carriers from being immediately confined by electrostatic potentials \cite{Katsnelson}, which limits the uses of graphene in electronic devices. For this reason, different ways of confining the quasiparticles in graphene has been proposed \cite{SiC,bdg2,BN,doped,bdg}, which includes the use of magnetic barriers \cite{PhysRevLett.98.066802,PhysRevB.77.245401,DEMARTINO2007547,MYOUNG200970,Milpas_2011,LE201817}. These barriers are the result of inhomogeneous magnetic fields that can be created, for instance, using ferromagnetic layers. With magnetic barriers it is possible, among other things, create bound states \cite{Ramezani}, control the transport properties \cite{Luca} and also break the valley degeneracy in graphene \cite{PhysRevB.86.115431,PhysRevB.85.155415}.

In the last years, several works have investigated the influence of a Fermi velocity modulation in the electronic and transport properties of graphene \cite{Ratnikov2008,LIMA201582,Krstaji__2011,LIMA20151372,Cheraghchi_2013,Ratnikov2014,Lima2017,ESMAILPOUR20121896,doi:10.1063/1.4953865,ARAUJO20173228,NASCIMENTO20192416}. For instance, it was obtained that the Fermi velocity can tune the energy gap \cite{LIMA2015179} and the Fano factor \cite{LIMA2018105} and also create electrons guides \cite{Polini,Yuan} and bound states in graphene \cite{Ghosh2017}. It was also showed that a Fermi velocity modulation can control the spin \cite{SATTARI2017438} and valley \cite{LINS2020353} transport in graphene. The Fermi velocity in graphene can be modulated, for instance, by doping \cite{doi:10.1002/pssb.200982335}, by the substrate \cite{Hwang}, by strain \cite{PhysRevB.84.195404,JANG2014139}, by electric fields \cite{Diaz-Fernandez2017} and also by placing metallic planes close to graphene \cite{Polini,Yuan}, which will change the charge concentration in different regions, inducing Fermi velocity barriers.  

The combination of magnetic barriers and a Fermi velocity modulation was already investigated. However, it is important to mention that it was considered only the cases of a single \cite{YUAN20114214} and double barrier \cite{LIU20123342}. Also, the barrier profile used in these works is different from what we will use here.

In this work we investigate the effects of a modulation of the Fermi velocity in the electronic and transport properties of magnetic graphene supeprlattices. We consider three cases: magnetic barriers, magnetic barriers and wells and a periodic superlattice. We use the transfer matrix method to obtain the transmission coefficient for the finite cases and the dispersion relation for the periodic case. We obtain, for instance, that the Fermi velocity modulation can control the transport properties, turning on/off the transmission through the magnetic barriers, which can be used for the fabrication of graphene-based electronic devices. We also find that the Fermi velocity can enhance or reduce the transmittance of the system and modulate the electronic structure.

The paper is organized as follows. In Sec. II we write out the effective Dirac equation of the system and the transmission coefficient for the finite cases and the dispersion relation for the periodic case. In Sec. III we obtain the results numerically and discuss the influence of the Fermi velocity in the electronic properties of the system. The paper is summarized and concluded in Sec. IV. 

\section{Model}

The effective Dirac equation that describes quasiparticles in a single layer graphene with a perpendicular magnetic field $B(x)$ and a modulated Fermi velocity $v_F(x)$ is given by

\begin{eqnarray}
\sqrt{v_F(x)}\vec{\sigma}\cdot \left(\vec{P} + \frac{e}{c}\vec{A}\right)\sqrt{v_F(x)}\Psi(x,y)=E\Psi(x,y),
\label{H}
\end{eqnarray}
where $\sigma = (\sigma_{x},\sigma_{y})$ are the Pauli matrices, $\vec{A}$ is the vector potential and $\Psi(x,y)$ is a spinor of two components that represents the two graphene sublattices. The Eq. (\ref{H}) was modified in relation to its usual form to become Hermitian, which is a consequence of the position dependence on the Fermi velocity \cite{peres}.

In the Landau gauge we have that $\vec{A} = (0,A_y(x),0)$. The wave functions are translationally invariants in the $y$ direction, which allows us to write $\Psi(x,y)=\psi(x)e^{ik_yy}$. Also, defining $\phi(x)=\sqrt{v_F(x)}\psi(x)$, Eq. (\ref{H}) becomes 
\begin{eqnarray}
\left[-i\partial_x \sigma_x + (k_y + A_y(x))\sigma_y \right]\phi(x) = \frac{E}{v_F(x) \hbar}\phi(x).
\label{eq}
\end{eqnarray}

We will use dimensionless units and express all quantities in units of $B_0$, which is the magnitude of the magnetic field, and $\ell_B=\sqrt{\hbar c/eB_0}$, the associated magnetic length. Then, $A_y(x)$ will be written in units of $B_0\ell_B$, $x$ in units of $\ell_B$, $k_y$ in units of $\ell_B^{-1}$ and $E$ in units of $\hbar v_F / \ell_B$. In this dimensionless units, the Fermi velocity modulation will be included in the problem by the energy. So, Eq. (\ref{eq}) give rises to two coupled equations given by
\begin{equation}
-i[\partial_x + (k_y + A_y)]\phi_B = E\phi_{A}
\end{equation}
and
\begin{equation}
-i[\partial_x - (k_y + A_y)]\phi_{A} = E\phi_{B}.
\label{ceq}
\end{equation}
Uncoupling these equations for $\phi_A$, one obtains that
\begin{equation}
\partial_{x}^2\phi_{A}+k^2_x (x) \phi_{A}=0,
\label{eqA}
\end{equation}
where $k_x(x)=\sqrt{E^2 - \partial_{x}A_y-(k_{y} + A_y)^2}$. The solution of this equation depends on the form of $A_y$. $\phi_B$ is obtained by replacing $\phi_A$ in Eq. (\ref{ceq}).

We will consider here three cases: $N$ magnetic barriers, $N$ magnetic barriers and wells and a periodic superlattice. 

\subsection{Magnetic barriers}

\begin{figure}[hpt]
\centering
\includegraphics[width=0.9\linewidth]{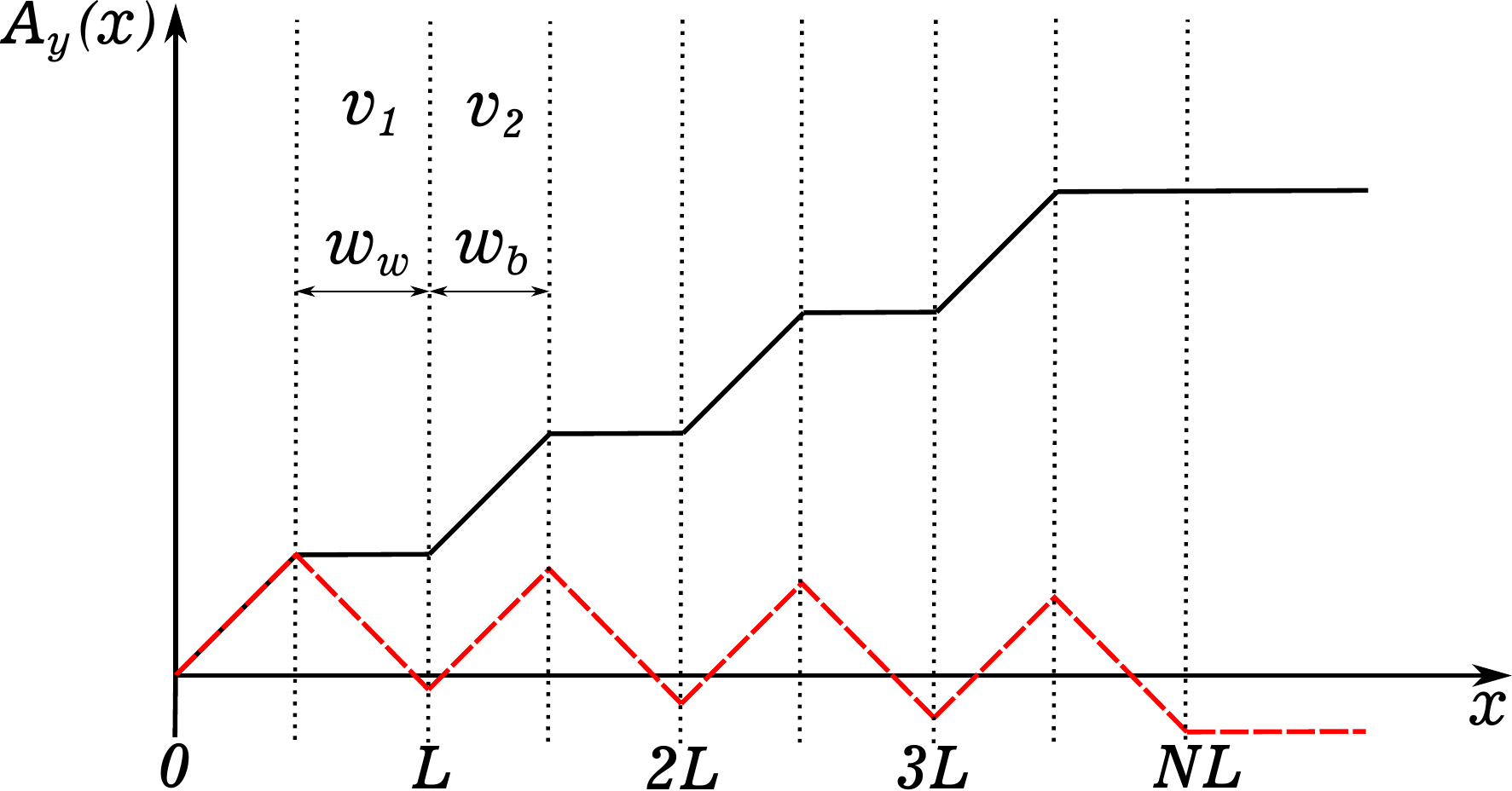}	
\caption{The $y$ component of the vector potential for the cases of magnetic barriers (continuum black lines) and magnetic barriers and wells (dashed red lines).}
\label{Ay}
\end{figure}

Let us first consider the case of $N$ magnetic barriers, which is shown in the continuum black lines of Fig. \ref{Ay}. We consider here that in the barrier regions the magnetic field and the Fermi velocity are given, respectively, by $\vec{B}=B_0\hat{z}$ and  $v_F=v_2$, while in the other regions $\vec{B}=0$ and $v_F=v_1$. The vector potential is given by

\begin{equation}
A_y(x) =  \left\{\begin{array}{cc}
0, & x\in [-\infty,0] \\ 
B_{0}(x-nw_w), & x\in [nL, nL+w_{b}] \\ 
B_{0}(n+1)w_{b}, & x\in [nL+w_b,(n+1)L] \\ 
Nw_{b}B_{0}, & x\in[NL, \infty]
\end{array}\right.
\end{equation}
where $n=0,..,N-1$ and $L=w_b+w_w$. 

In the regions with constant $A_y$, we have that
\begin{equation}
    \phi_0(x)=A_0\left(
    \begin{array}{c}
      1 \\
     \frac{k_x+i(k_y+A_y)}{E}
    \end{array}\right)e^{ik_xx}+C_0\left(
    \begin{array}{c}
         1  \\
         \frac{-k_x+i(k_y+A_y)}{E} 
    \end{array}\right)e^{-ik_xx},
\end{equation}
where $A_0$ and $C_0$ are constants that give the amplitude of the waves in the region.
This solution can be written as
\begin{equation}
    \phi_0(x)=\Omega_0\left( 
    \begin{array}{c}
         A_0  \\
         C_0 
    \end{array}\right),
\end{equation}
where
\begin{equation}
\Omega_{0} = \left(\begin{array}{cc}
e^{ik_{x}x} & e^{-ik_{x}x}\\ 
\frac{k_{x}+i(k_{y}+A_y)}{E}e^{ik_{x}x} & \frac{-k_{x}+i(k_{y}+A_y)}{E}e^{-ik_{x}x}
\end{array}\right).
\end{equation}

For the regions where $A_y$ depends on position, we have that
\begin{equation}
    \phi_b(x)=\Omega_b\left( 
    \begin{array}{c}
         A_b  \\
         C_b 
    \end{array}\right),
\end{equation}
where
\begin{equation}
\Omega_{b} = \left(\begin{array}{cc}
D_{p}(q) & D_{p}(-q)\\ 
\frac{i\sqrt{2B_0}}{E\zeta}D_{p+1}(q)& -\frac{i\sqrt{2B_0}}{E\zeta}D_{p+1}(-q)
\end{array}\right),
\end{equation}
with $q=\sqrt{2/B_0}(k_y+A_y(x))$, $\zeta=v_2/v_1$ and $p=(E\zeta)^2/(2B_0)-1$ and $D_p(q)$ is the parabolic cylinder function. 

Considering the continuity condition for $\phi(x)$ in the interface of each region, it is possible to obtain a matrix that connects the amplitude of the waves in the incoming and outgoing regions, which is given by
\begin{equation}
 \left(\begin{array}{cc}A_i \\ C_i \end{array}\right) =\hat{M} \left(\begin{array}{cc}A_f \\ C_f \end{array}\right),
 \label{tm}
\end{equation} 
where the indices $i$ and $f$ represent, respectively, the incoming and outgoing regions and
\begin{equation}
    \hat{M}=\hat{M}_0\hat{M}_1...\hat{M}_{N-1}
\end{equation}
with
\begin{equation}
\hat{M}_n = \Omega_{0}^{-1}(nL) \Omega_{b}(nL)\Omega_{b}^{-1}(nL+w_b) \Omega_{0}(nL+w_b).
\end{equation}
It is important to remember that the wave function of the problem is $\psi(x)$, and not $\phi(x)$. However, since we are considering that the Fermi velocity is the same in the incoming and outgoing regions, the matrix $\hat{M}$ obtained for $\phi$ is the same for $\psi$.

From Eq. (\ref{tm}) one can obtain that the transmittance is given by
\begin{equation}
    T=\frac{k_x^f}{k_x^i} \frac{|A_f|^2}{|A_i|^2}=\frac{k_x^f}{k_x^i} \frac{1}{|M_{11}|^2},
\end{equation}
where $k_x^i$ and $k_x^f$ are the $x$ component of the wave vector in the incoming and outgoing regions, respectively, and we put $C_f=0$.

Writing the wave vectors in terms of the incident and exit angles, $\theta_0$ and $\theta_f$, respectively, we have that
\begin{equation}
    k_{x}^i = E\cos \theta_0,\;\; k_{y}^i = E\sin \theta_0
\end{equation}
\begin{equation}
    k_{x}^f = E\cos \theta_{f},\;\; k_{y}^f = E\sin \theta_{f} - Nw_b B_0,
\end{equation}
where $Nw_bB_0$ is the total magnetic flux. The conservation of $k_y$ implies that $k_y^i=k_y^f$, which gives us
\begin{equation}
    \sin \theta_0 + \frac{N w_b B_0}{E} = \sin \theta_f.
\end{equation}
So, the transmission through the barriers is only possible when
\begin{equation}
   \left| \sin \theta_0 + \frac{N w_b B_0}{E}\right| \leq 1,
   \label{anglelimit}
\end{equation}
which restricts the transmission for a smaller range of $\theta_0$ then $[-\pi/2, \pi/2]$.

\subsection{Magnetic barriers and wells}

For the case of magnetic barriers and wells, the vector potential is given by
\begin{equation}
A_y(x) =  \left\{\begin{array}{cc}
0, & x\in [-\infty,0] \\ 
B_{0}(x-2nw_w), & x\in [nL, nL+w_{b}] \\ 
B_{0}[2(n+1)w_{b}-x], & x\in [nL+w_b,(n+1)L] \\ 
N(w_{b}-w_w)B_{0}, & x\in[NL, \infty]
\end{array}\right.
\end{equation}
This vector potential can be seen in the red dashed lines in Fig. \ref{Ay}. Here, we have that $\vec{B}=B_0\hat{z}$ and $v_F=v_2$ in the barrier regions and $\vec{B}=-B_0\hat{z}$ and $v_F=v_1$ in the well regions.

Following the same procedure as in the previous section, one can obtain that the solution in the well regions is given by
\begin{equation}
    \phi_w(x)=\Omega_w\left( 
    \begin{array}{c}
         A_w  \\
         C_w 
    \end{array}\right),
\end{equation}
where
\begin{equation}
\Omega_{w} = \left(\begin{array}{cc}
D_{p+1}(-q) & D_{p+1}(q)\\ 
\frac{i\sqrt{2B_0}}{E\zeta}(p+1)D_{p}(-q)& -\frac{i\sqrt{2B_0}}{E\zeta}(p+1)D_{p}(q)
\end{array}\right),
\end{equation}
and the transfer matrix is 
\begin{equation}
    \hat{M}=\Omega_0^{-1}(0)\Omega_w(0)\hat{M}_0\hat{M}_1...\hat{M}_{N-1}\Omega_w^{-1}(NL)\Omega_0(NL),
\end{equation}
where here
\begin{equation}
\hat{M}_n = \Omega_{w}^{-1}(nL) \Omega_{b}(nL)\Omega_{b}^{-1}(nL+w_b) \Omega_{w}(nL+w_b).
\end{equation}
In this case, the total flux is given by $N(w_b-w_w)B_0$. So, we have that
\begin{equation}
   \left| \sin \theta_0 + \frac{N (w_b-w_w) B_0}{E}\right| \leq 1,
   \label{anglelimit2}
\end{equation}

\subsection{Periodic Superlattice}

Here we consider the case of a periodic magnetic superlattice, which is possible considering barriers and wells with $w_b=w_w$. From the transfer matrix method, we have that
\begin{equation}
    \psi(0)=\Omega \psi(L),
    \label{tma}
\end{equation}
where
\begin{equation}
    \Omega= \Omega_{b}^{-1}(0) \Omega_{w}(0)\Omega_{w}^{-1}(w_b)\Omega_b(w_b).
\end{equation}
The Bloch theorem says that
\begin{equation}
    \psi(0)=e^{-iK_xL}\psi(L),
    \label{bloch}
\end{equation}
where $K_x$ is the Bloch wave number. Comparing Eqs. (\ref{tma}) and (\ref{bloch}), it is possible to see that
\begin{equation}
    \det[\Omega-\exp(-iK_xL)]=0,
\end{equation}
which yields the dispersion relation
\begin{equation}
    2 \cos K_xL = Tr [\Omega]. 
\end{equation}

\section{Numerical results and discussions}

With the transmission coefficient for the finite cases and the dispersion relation for the periodic case, we can now obtain and analyze the results. We will consider each case separately.

\subsection{Magnetic barriers}

\begin{figure}[hpt]
\centering
\includegraphics[width=0.9\linewidth]{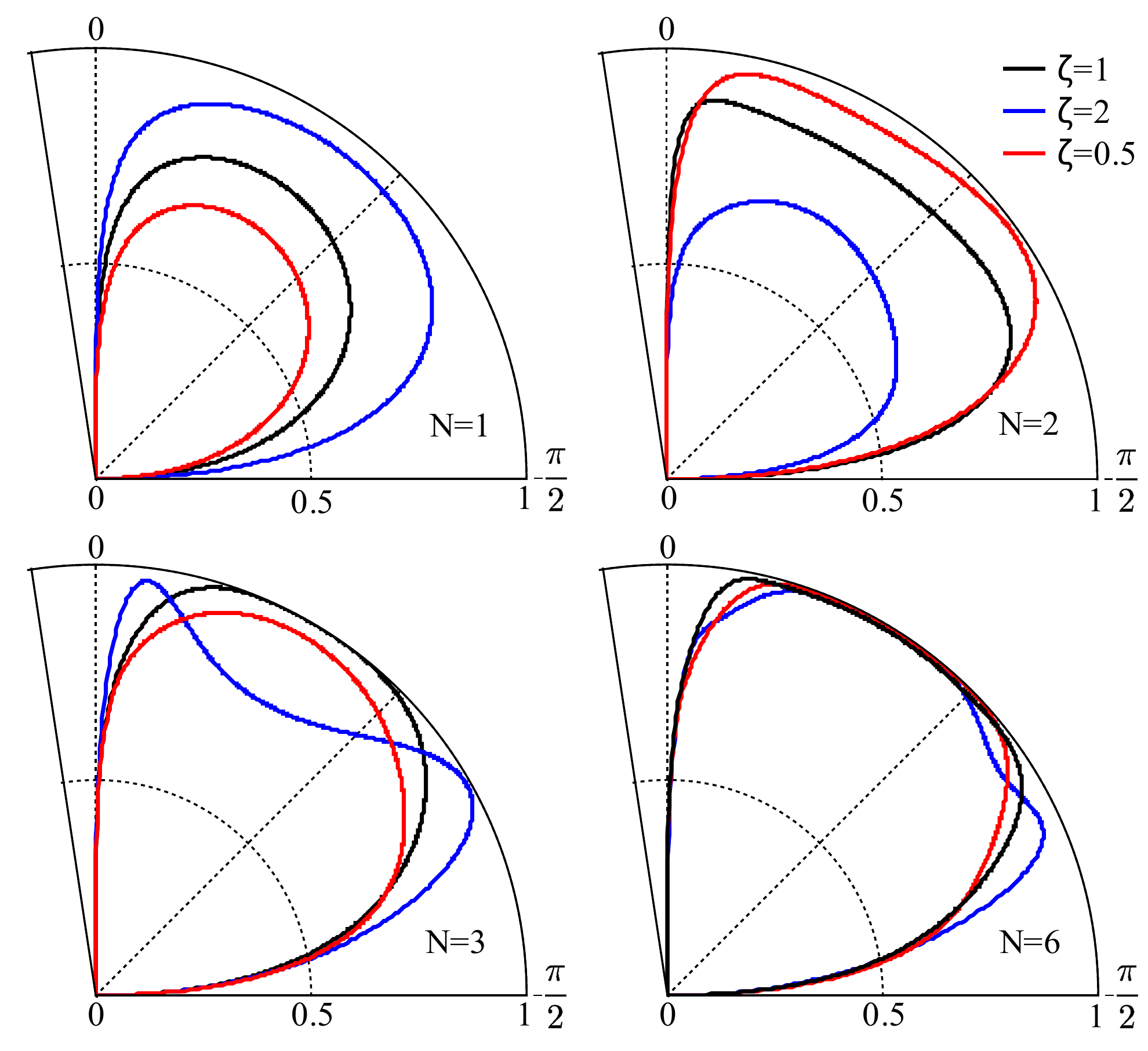}	
\caption{The transmittance as a function of the incidence angle for different values of N and $\zeta$. We are considering here $B_0=E=1$, $w_w=5$ and $w_b=1/N$.}
\label{N-B}
\end{figure}

In Fig. \ref{N-B} we plotted the transmittance as a function of the incidence angle for different values of $N$ keeping the total magnetic flux constant.  We consider the cases with $\zeta$ equal to $0.5$, $1$ and $2$. As expected, the range of $\theta_0$ with transmission different from $0$ does not change, since the flux is the same in all plots. 

For the case without a Fermi velocity modulation (black lines) one can see that, as the magnetic flux is divided by adding more barriers, the transmission increases and tends to the classical limit, where the transmission is 1 when Eq. (\ref{anglelimit}) is satisfied or 0 otherwise. However, with a modulation of the Fermi velocity, the transmission can also decreases as the number of barriers increases, as occur for $\zeta=2$ when $N$ goes from 1 to 2 and for $\zeta=0.5$ when $N$ increases from 2 to 3. But, for a large value of $N$, all cases approach the classical limit, as can be seen for $N=6$. So, a change in $\zeta$ can increase or decrease the transmission depending on $N$. One can also see that for $\zeta=2$ some oscillations in the transmission appear for high values of N.

\begin{figure}[hpt]
\centering
\includegraphics[width=0.9\linewidth]{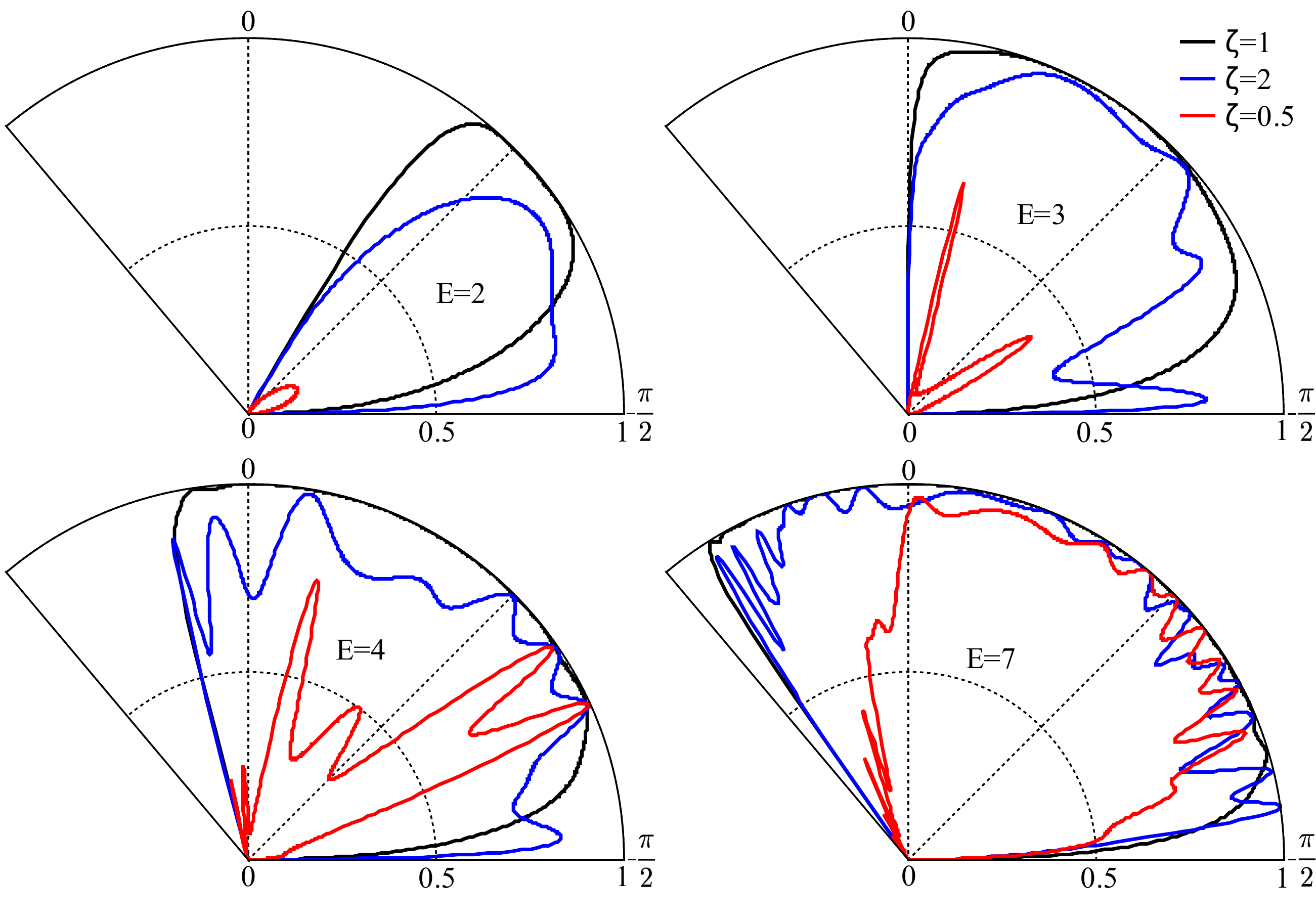}	
\caption{The transmittance as a function of the incidence angle for different values of energy and $\zeta$. The other parameters are: $N=3$, $B_0=w_b=1$ and $w_w=5$.}
\label{E-B}
\end{figure}

The transmittance for various values of energy is shown in Fig. \ref{E-B}. As can be seen, the range of $\theta_0$ with non-null transmission increases with $E$, in accordance with Eq. (\ref{anglelimit}). In the case with $E=2$ it is clear that, depending on the value of $\zeta$, the trnasmittance can be enhanced or suppressed, revealing that the Fermi velocity modulation can control the transport properties in these magnetic graphene superlattices.

One can also note the appearance of oscillations in the transmission for higher values of $E$. It is possible to thought these magnetic barriers as Fabry-P\'erot interferometers, where different incidence angles give rise to constructive or destructive interferences. At this way, the modulation of the Fermi velocity can control the Fabry-P\'erot resonances, selecting which incidence angle will be transmitted,  and also cancel the interference process, as occurs for $\zeta=1$.

\begin{figure}[hpt]
\centering
\includegraphics[width=\linewidth]{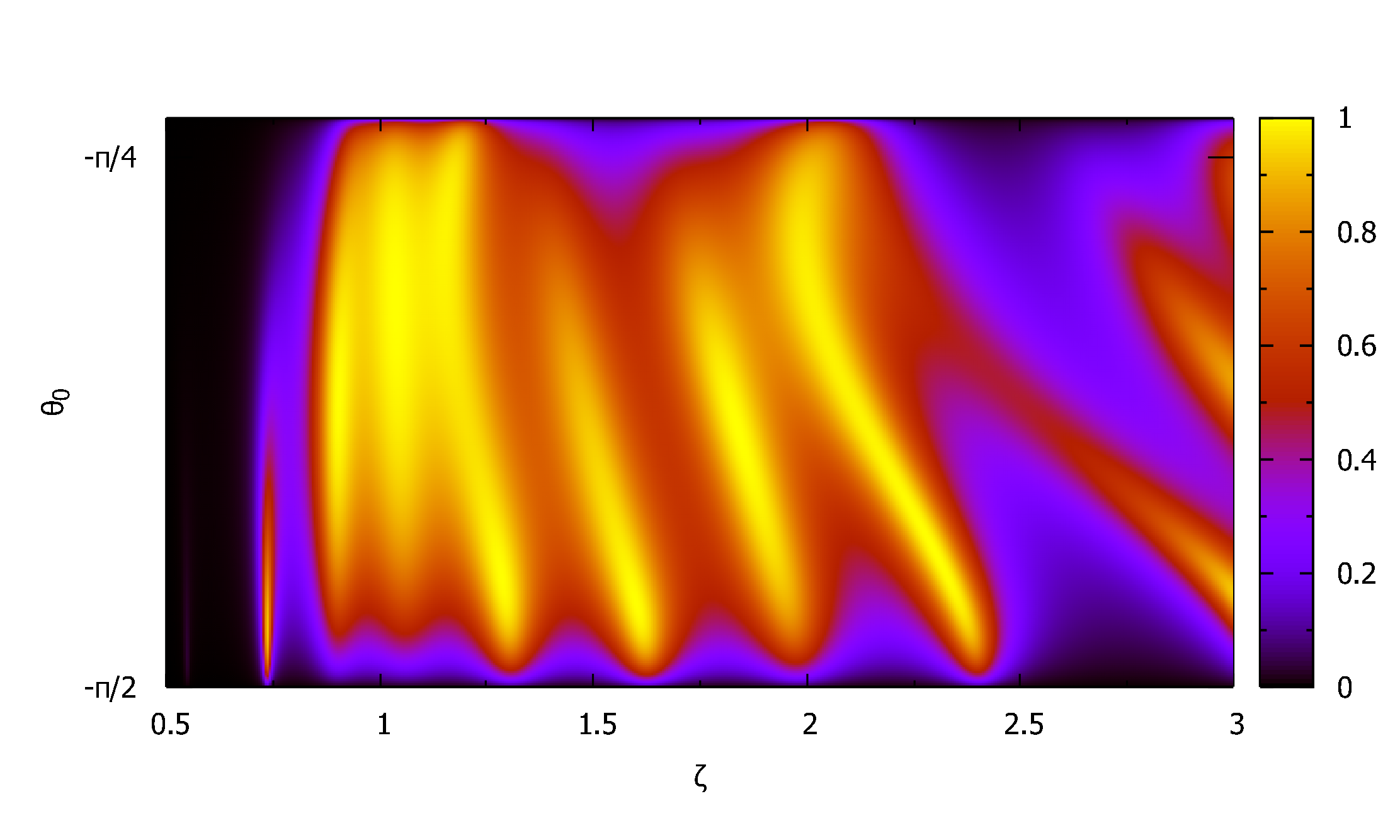}	
\caption{Contour plot of the transmittance as a function of the incidence angle and $\zeta$ with $B_0=w_b=w_w=1$, $N=5$ and $E=3$.}
\label{c-B}
\end{figure}

The control of the transmittance by a modulation of $v_F$ can be seen more clearly in Fig. \ref{c-B}, where we have a contour plot of the transmittance in terms of the incidence angle and $\zeta$. We can see that the transmission for each incidence angle oscillates with a change in the value of $\zeta$.

\subsection{Magnetic barriers and wells}

\begin{figure}[hpt]
\centering
\includegraphics[width=0.9\linewidth]{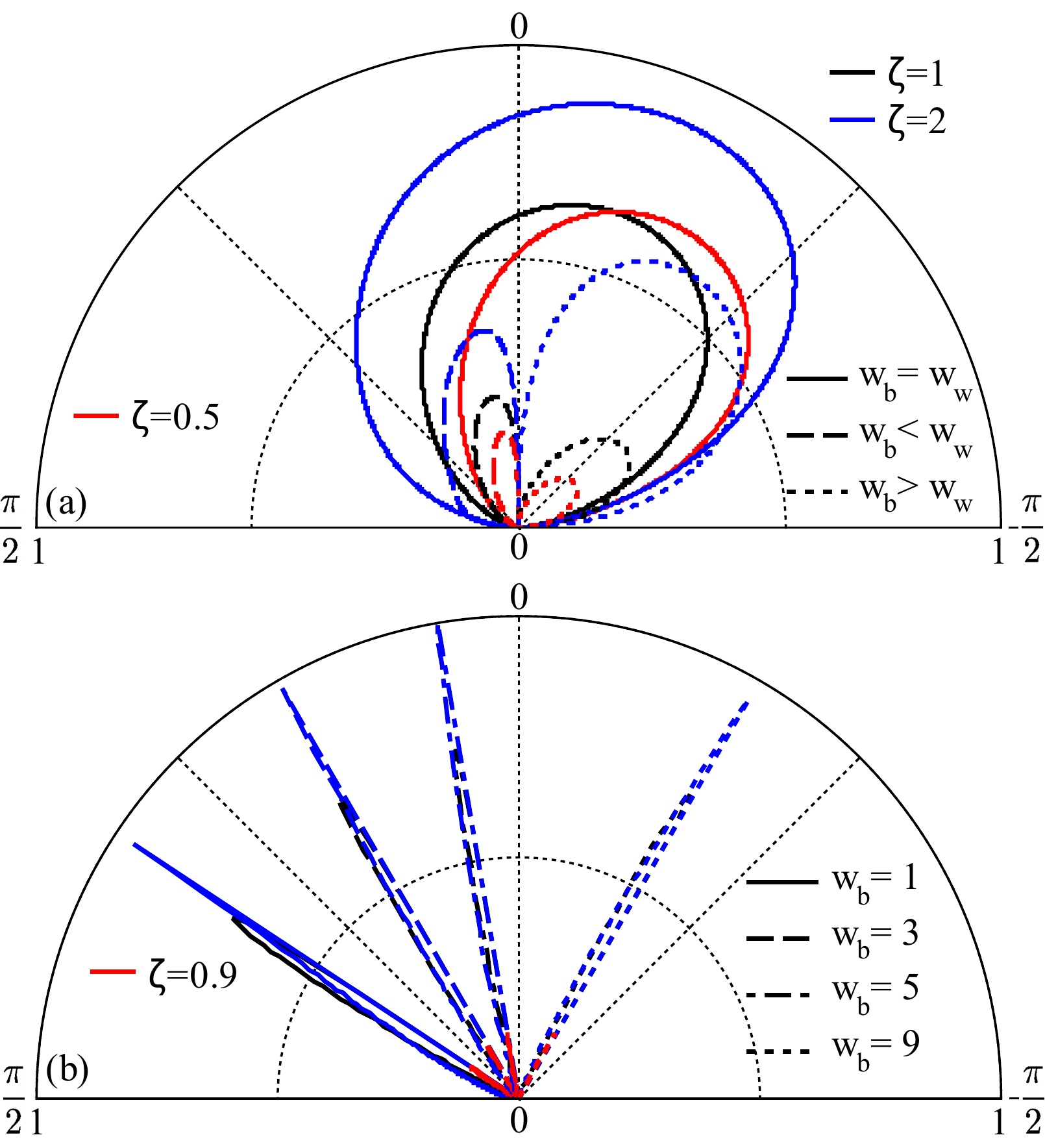}	
\caption{The transmittance as a function of the incidence angle for different values of $w_b$, $w_w$ and $\zeta$. We consider here $N=E=B_0=1$. (a) $w_b=w_w=1$ (continuum lines), $w_b=1$ and $w_w=2$ (dashed lines), $w_b=2$ and $w_w=1$ (dotted lines). (b) $w_w=6$.}
\label{w-BW}
\end{figure}

Let us now consider the case with magnetic barriers and wells. In Fig. \ref{w-BW} it is possible to see the influence of the Fermi velocity modulation in the transmittance for different values of the widths of the regions. In Fig. \ref{w-BW} (a) one can see that when $w_b<w_w$ $(w_b>w_w)$ the transmittance is different from zero only for positive (negative) incidence angles. It can be understood looking to the sign of the total flux in Eq. (\ref{anglelimit2}). For $w_b=w_w$ the total flux in equal to zero, which means that the transmission is not limited to a smaller range of the incidence angle. In all cases an increase in $\zeta$ induces an enhancement on the transmittance, while a decrease in $\zeta$ reduces the transmission.

Increasing the width of the regions, it is possible to restrict the transmission to a very small range of incidence angle, as can be seen in Fig. \ref{w-BW} (b), where there is only a very narrow transmission peak for a specific incidence angle. We are considering here $w_w=6$ and the value of $w_b$ determines the incidence angle that can be transmitted, which, consequently, aligns the transmitted electrons in a specific direction. So, such system can be used as a collimator of electrons beams, which would be very useful, for instance, in experiments with graphene superlattices, since it is not easy to control the direction of propagation of the quaseparticles in such systems. As can be seen, a Fermi velocity modulation can not change the location of the transmission peak, but it can tune the transmittance from 0 to 1. So, it can be used as a switch, turning on/off the transmission in the system.

\begin{figure}[hpt]
\centering
\includegraphics[width=0.9\linewidth]{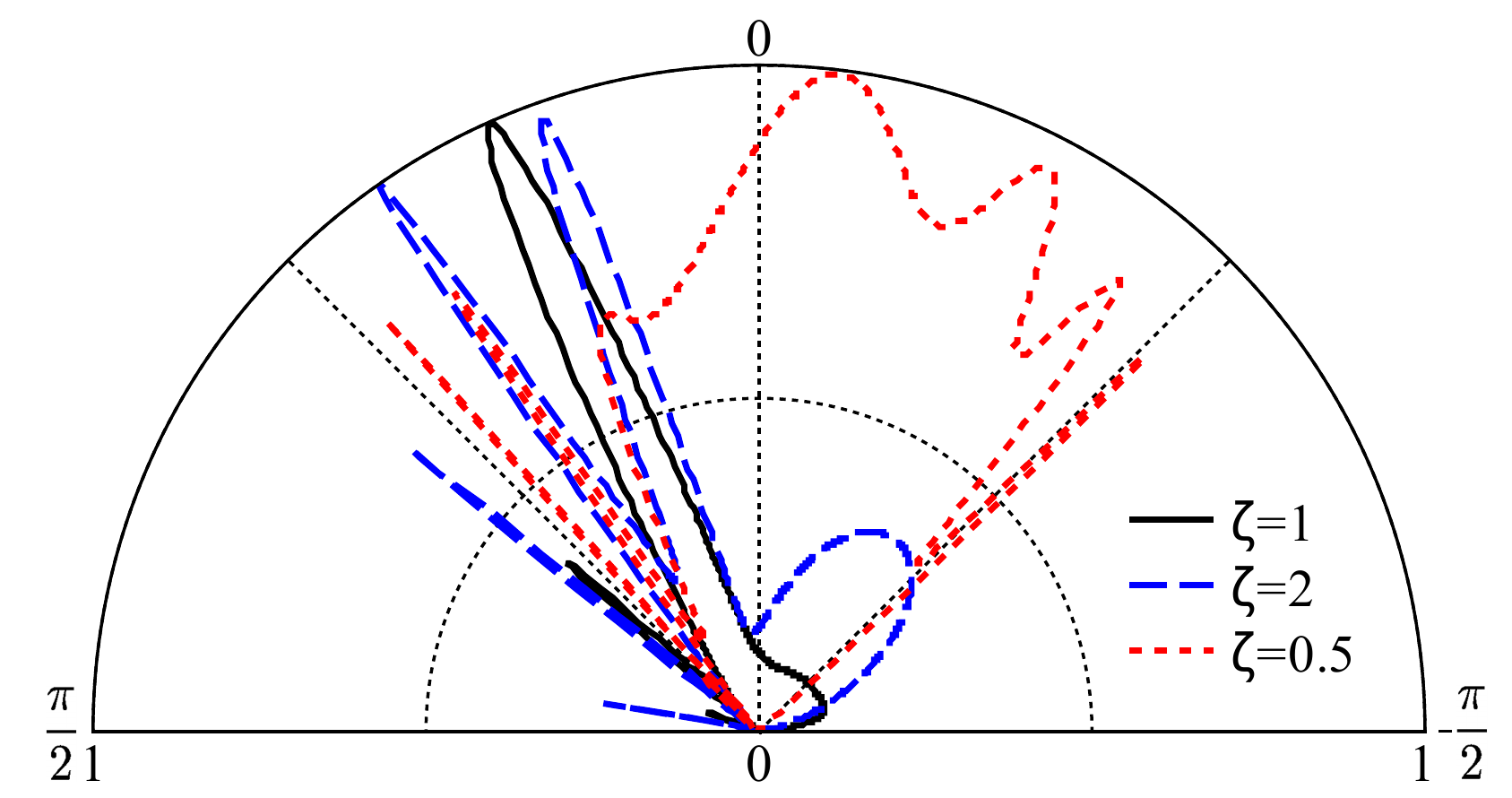}	
\caption{The transmittance as a function of the incidence angle for different values of $\zeta$. We consider here $B_0=w_b=w_w=1$, $N=6$ and $E=3$.}
\label{bwn6}
\end{figure}

In Fig. \ref{bwn6} we consider the case with $N=6$. One can see here the appearance of oscillations in the transmittance and also narrow peaks in the transmittance for some values of the incidence angle, which are the consequence of the interference process. The Fermi velocity modulation can control these interference, since the resonance peaks occur for different incidence angle as we change the value of $\zeta$.  

\begin{figure}[hpt]
\centering
\includegraphics[width=\linewidth]{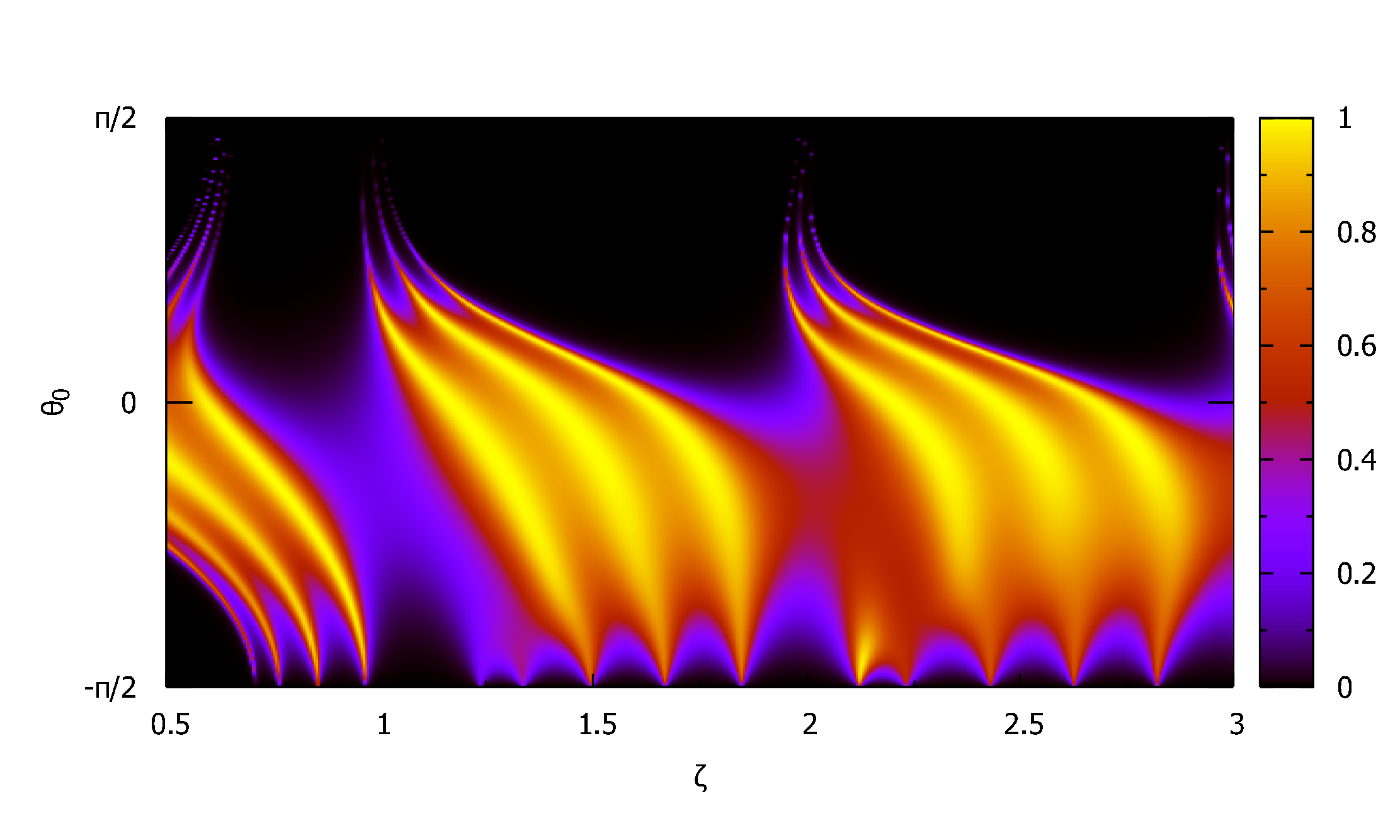}	
\caption{Contour plot of the transmittance as a function of the incidence angle and $\zeta$ with $B_0=w_b=w_w=1$, $N=5$ and $E=3$.}
\label{c-BW}
\end{figure}

A contour plot of the transmittance as a function of the incidence angle and $\zeta$ can be seen in Fig. \ref{c-BW}. As can be see, the transmittance oscillates with $\zeta$, revealing the control of the transmission by the Fermi velocity. 

\subsection{Periodic superlattice}

\begin{figure}[hpt]
\centering
\includegraphics[width=\linewidth]{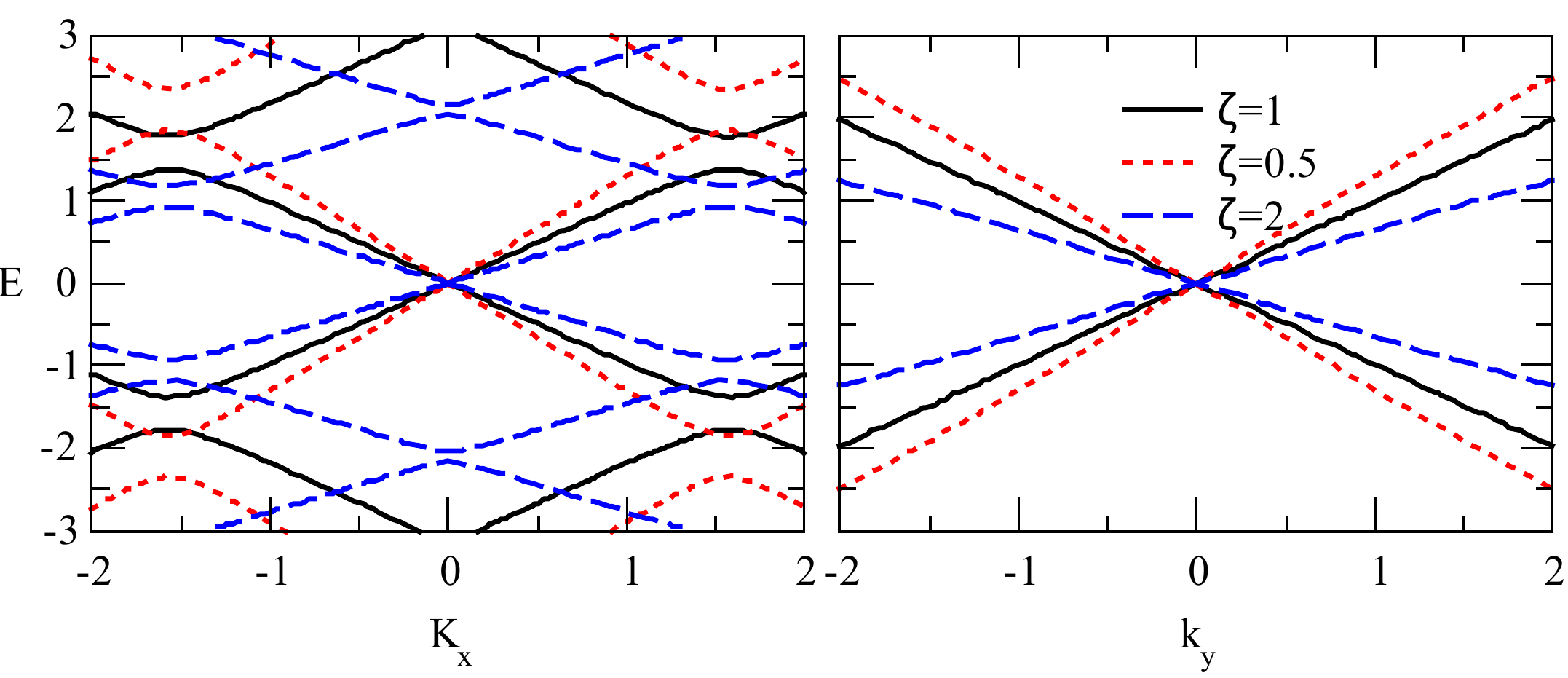}	
\caption{Dispersion relation for the periodic superlattice. We consider the energy in terms of $K_x$ for $k_y=0$ (left panel) and the energy as a function of $k_y$ with $K_x=0$ (right panel). We have here that $B_0=w_b=w_w=1$.}
\label{kxky}
\end{figure}

The band structure for the periodic superlattice case can be seen in Fig. \ref{kxky}. We can see a linear dispersion relation, with a conical surface similar to the graphene without the magnetic field. However, the dispersion relation here is inversely proportional to $\zeta$, since an increase (decrease) in $\zeta$ also increases (decrease) the angle of the conical surface. 

\section{Conclusion} 

In this paper we have investigated the influence of a Fermi velocity modulation in the electronic and transport properties of magnetic graphene superlattices. We obtained that, for the case of finite magnetic barriers, the Fermi velocity can enhance or reduce the transmission through the magnetic barriers and also to control the resonant peaks in the transmittance. We also showed that the magnetic field can select the incidence angle that will be transmitted, which can be used to create a colimator of electrons beams. The Fermi velocity in such colimator works as a switch, since it can tune the transmittance from 0 to 1. For the case of a periodic magnetic graphene superlattice, we found a linear dispersion relation that is proportional to the ratio of the Fermi velocity in the two regions of the superlattice. The results obtained here are useful in the development of new electronic devices based on graphene. 

{\bf Acknowledgements}: This work was supported by CNPq, Capes and Alexander von Humboldt Foundation.

\end{document}